\documentclass[final,1p,times]{elsarticle} 
\usepackage{graphicx}
\usepackage{amssymb} 
\usepackage{amsthm}

\begin{document}

\begin{frontmatter} 
\title{The statistical model in Pb-Pb collisions at the LHC}

\author[a1]{A. Andronic}
\author[a1,a2,a3]{P. Braun-Munzinger}
\author[a4]{K. Redlich}
\author[a5]{J. Stachel}
\address[a1]{Research Division and ExtreMe Matter Institute EMMI and , GSI 
Helmholtzzentrum f\"ur Schwerionenforschung, Darmstadt, Germany}
\address[a2]{Technical University, Darmstadt, Germany}
\address[a3]{Frankfurt Institute for Advanced Studies, J.W. Goethe University,
Frankfurt, Germany}
\address[a4]{Institute of Theoretical Physics, University of Wroc\l aw, Poland}
\address[a5]{Physikalisches Institut, University of Heidelberg, Germany}

\begin{abstract}
We briefly review the predictions of the thermal model for hadron
production in comparison to latest data from RHIC and extrapolate the
calculations to LHC energy. Our main emphasis is to confront the model
predictions with the recently released data from ALICE at the
LHC. This comparison reveals an apparent anomaly for protons and
anti-protons which we discuss briefly. We also demonstrate that our
statistical hadronization predictions for J/$\psi$ production agree
very well with the most recent LHC data, lending support to the
picture in which there is complete charmonium melting in the
quark-gluon plasma (QGP)  
followed by statistical generation of J/$\psi$ mesons at the phase boundary.  
\end{abstract}

\end{frontmatter} 

The quark-gluon plasma produced in ultra-relativistic nuclear
collisions profoundly influences the production of hadrons 
\cite{pbm_wambach,pbm_js,wetterich}.  For central collisions between
large nuclei the yields of hadrons made up of light quarks can be
described very well from AGS to RHIC energies (see \cite{aat} and
refs. therein) within the statistical model, with the
chemical freeze-out temperature $T$, the baryo-chemical potential
$\mu_b$ and the fireball volume $V$ as the only parameters.  
Since the
extracted temperature values, which first increase sharply with
increasing beam energy, level off near $T$=160 MeV for energies
$\sqrt{s_{NN}}>$20 GeV, while the baryochemical potential decreases
smoothly as a function of energy, the extrapolation to LHC energy is
straightforward. Consequently, analysis of the recently released
hadron production data from ALICE at the LHC provides a crucial test
for the statistical model. This was first investigated in
\cite{aa_qm11} for data on hadrons with light and heavy quarks. Here
we provide an update and compare to the most recent data. 

\begin{figure}[h]
\begin{center}
\includegraphics[width=.6\textwidth]{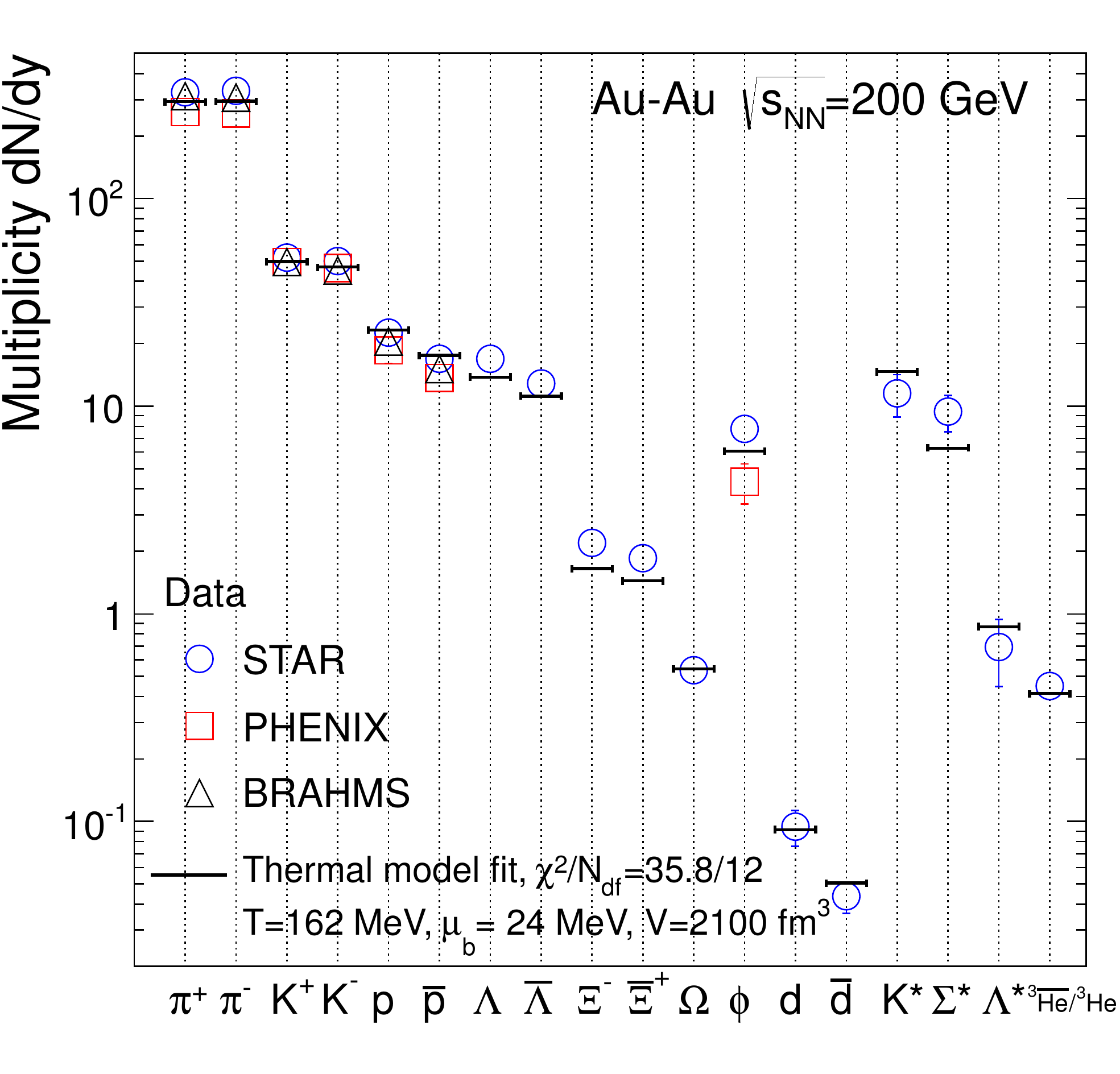}
\caption{Comparison of thermal model predictions with RHIC data.The
  data are as compiled in \cite{aat}, with a recent update taking into
account all available information on feeding via weak decays of
multi-strange baryons.}
\label{rhic1}
\end{center}
\end{figure}

Before proceeding to LHC data we show, in Fig.~\ref{rhic1},  the
situation for full RHIC energy ($\sqrt{s_{NN}}=$200 GeV). The
overall trend of the data is very well reproduced by the calculations.
However, the yield of protons and anti-protons from PHENIX and
Brahms is overpredicted, while the yield of multi-strange baryons is
generally underpredicted, leading to a rather poor $\chi^2$ value of
the fit\footnote{A fit to data from the STAR experiment alone leads to
  a significantly improved fit.}. We note that feeding from weak
decays of multi-strange baryons could be much improved with vertex
detectors, as have been (PHENIX) or will be (STAR) installed into the
RHIC detectors. Measurements with vertex detectors would be very
important to provide a more uniform experimental picture. 

\begin{figure}[htb]
\begin{tabular}{cc}
\begin{minipage}{.49\textwidth}
\includegraphics[width=.95\textwidth]{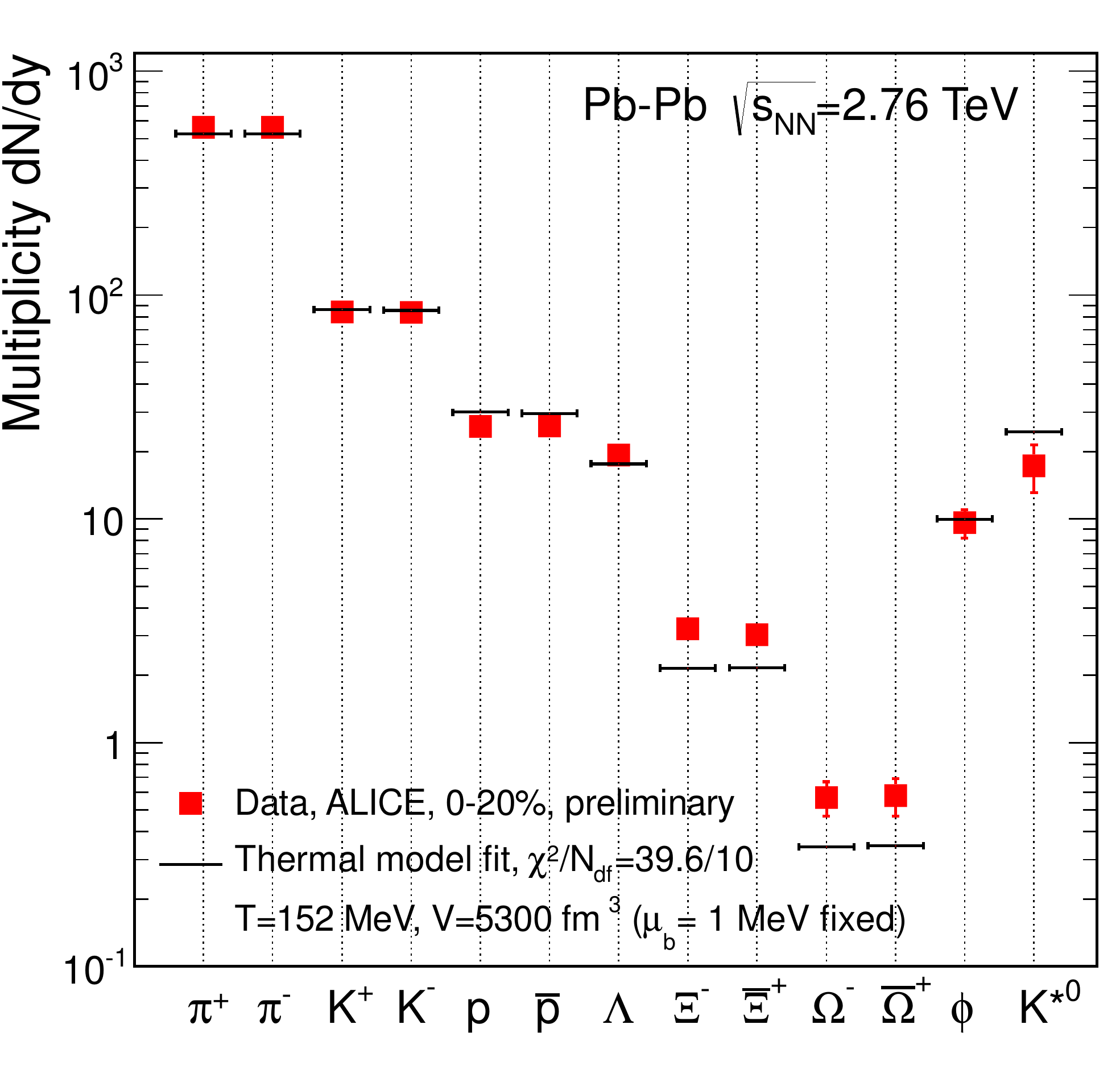}
\end{minipage}  & \begin{minipage}{.49\textwidth}
\includegraphics[width=.95\textwidth]{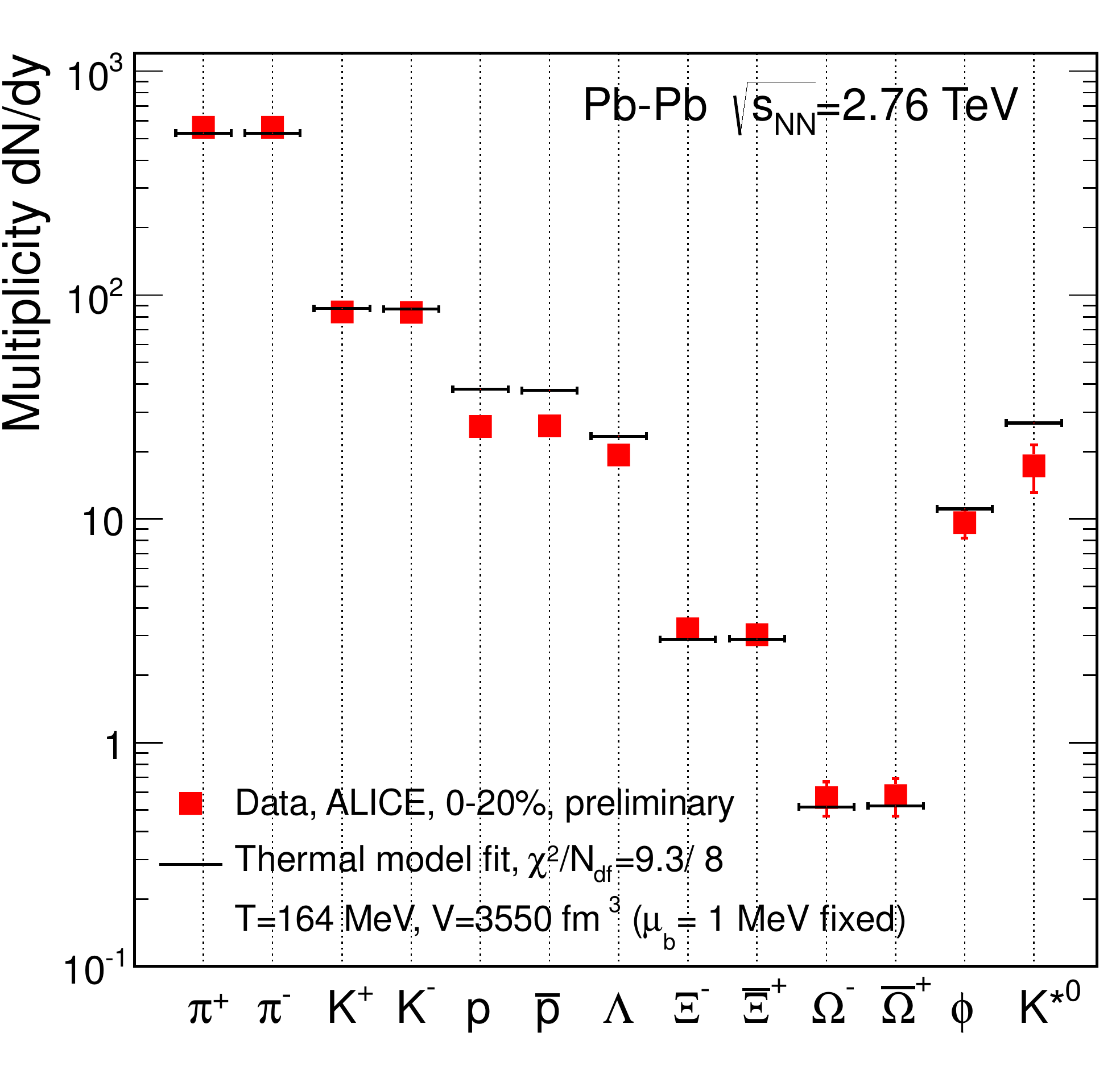}
\end{minipage}\end{tabular}
\caption{Thermal model fits to ALICE data on hadron production in
  central Pb--Pb collisions. The left panel shows the result of the
  fit to all available data, while protons and anti-protons are
  excluded from the fit shown in the right panel. 
The ALICE data are preliminary results shown at this 
conference \cite{alice_1}.}
\label{alice_had}
\end{figure}

In Fig.~\ref{alice_had} we show the results of thermal model fits to
reently released ALICE data \cite{alice_1}. The left panel shows a fit
to all currently available data. The unexpectedly low yields for
protons and anti-protons drive the temperature of the fit to a rather
low value ($T = 152$ MeV) while the yield of multi-strange baryons is
significantly underpredicted. This is somewhat similar to the
situation observed at RHIC (Fig.~\ref{rhic1}). With the more than
a factor of 2 smaller error bars of the ALICE data compared to results
from the RHIC experiments the reduced $\chi^2$ value approaches 4, and
the temperature parameter is significantly lower than expected from
the extrapolation from the data at lower energies \cite{aat}.

The right hand panel in Fig.~\ref{alice_had} shows the result of
excluding protons and anti-protons from the fit. This leads to a very
good description of all remaining data, with excellent $\chi^2$
parameter and a temperature value (164 MeV) completely in line with
expectations. Naturally, the nucleon yields are now about a factor of 1.4
below the calculated values. This apparent proton anomaly could be due
to annihilation in the hadronic phase near the phase boundary. Indeed,
schematic model calculations indicate such an effect
\cite{bleicher,pratt}. We note, however, that annihilation affects not
only nucleons, but also strange and multi-strange baryons. If
annihilation is the explanation for the proton anomaly then the new
ALICE data suggests that the annihilation rate for strange baryons is
significantly less than that for nucleons. Further precision
measurements, including also correlations among baryons and
anti-baryons, are needed to shed light on this observation.

In the following we use the statistical model to make predictions for
charmonium production and compare the results to the most recent ALICE
data \cite{alice_2a,alice_2b}.  Suppression of J/$\psi$ mesons in the QGP was
originally predicted \cite{satz} as a key signature for a dense
partonic phase. In contrast, in \cite{pbm1} it was argued that
charmonium production can be well described in the statistical model
by assuming that all charm quarks are produced in initial, hard
collisions. An important further input is that  the QGP provides
complete color screening, implying that charmed hadrons and charmonia are 
first produced at the phase boundary with statistical weights (for a recent 
review see \cite{pbm_js_lb}, for a detailed more technical description see
\cite{aa2}).  An important element is thermal equilibration of  
charm quarks, at least near the transition temperature $T_c$. The new
ALICE data \cite{alice_2d} on spectra and flow of open charm hadrons
and charmonia provide good evidence for this.  

\begin{figure}[htb]
\begin{tabular}{cc}
\begin{minipage}{.49\textwidth}
\includegraphics[width=.95\textwidth]{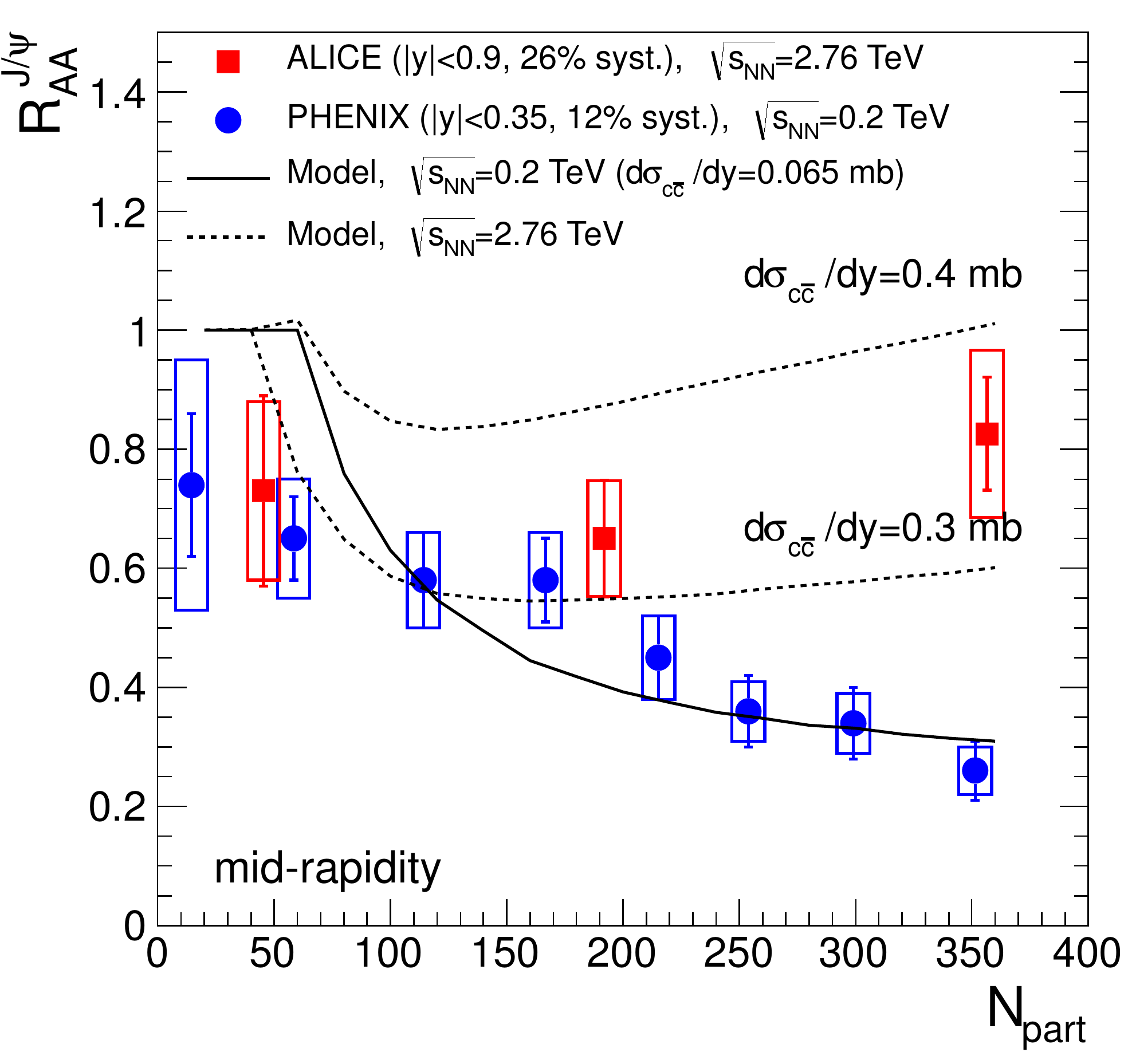}
\end{minipage}  & \begin{minipage}{.49\textwidth}
\includegraphics[width=.95\textwidth]{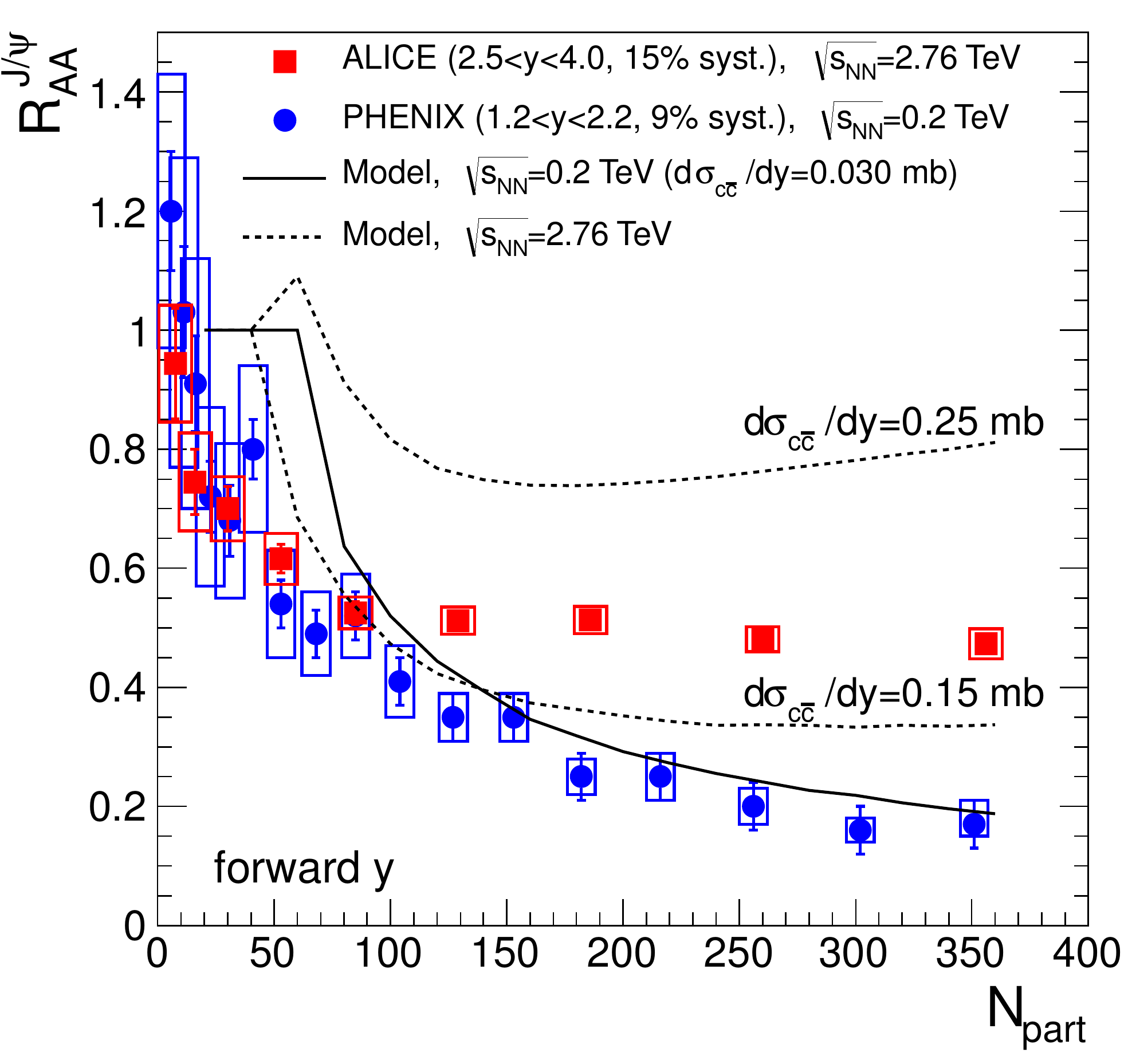}
\end{minipage}\end{tabular}
\caption{Centrality dependence of $R_{AA}^{J/\psi}$ for RHIC and LHC energies
at mid-rapidity (left panel) and forward rapidity (right panel). The two curves
shown for the LHC energy coorespond to a range of expected shadowing. 
The ALICE data shown in the left panel are preliminary results shown at this 
conference \cite{alice_2a}.}
\label{fig_raa}
\end{figure}

The centrality dependence of the nuclear modification factor
$R_{AA}^{J/\psi}$ as measured recently by ALICE \cite{alice_2b} is
shown in Fig.~\ref{fig_raa}, for central and forward rapidity, and
compared to RHIC data from the PHENIX collaboration \cite{phe1} as well as to
predictions from the statistical hadronization model. We first note
that, at LHC energy, much less suppression is observed compared to the
RHIC results, both at forward- and at mid-rapidity. The model
calculations \cite{aa2} 
reproduce this trend very well.  In our model the larger
$R_{AA}^{J/\psi}$ values at 
midrapidity are due to the enhanced generation of
charmonium around mid-rapidity, determined by the rapidity dependence
of the charm production cross section.  Also the observed centrality
dependence is correctly reproduced.

The successful description of the new ALICE data lends strong support
to the interpretation that, at LHC energy, J/$\psi$ mesons do not form
or survive inside the QGP, implying strong color screening. 
Rather, the observations are consistent with the formation of charmonium bound
states at hadronization of the QGP. 
Conceptually, this is very different from the mechanism of continuous formation
and destruction of charmonia in the QGP, as employed in transport models
\cite{rr,pfz}.
In our model, charmonium production is a direct signal for deconfinement 
of charm quarks: the charmonia are dominantly formed from initially 
uncorrelated $c$ and $\bar c$ quarks. 
Further measurements of $\psi'$ and $\chi_c$ states, as planned with the 
ALICE upgrade project \cite{alice_loi} will be crucial to differentiate 
between the models.
The next step is a measurement in pPb collisions, which will clarify the 
contribution of shadowing.
In Pb-Pb collisions measurements will be performed at full LHC energy.
Due to the increase of the charm production cross section, we expect a further 
increase of $R_{AA}^{J/\psi}$ of up to 40\% at mid-rapidity for central collisions
\cite{last_call}.

\end{document}